\documentclass[11pt]{article}

\usepackage{epsfig,picinpar,latexsym,amsmath,amsfonts}

\newcommand{\etal}{{\em et al.}}

\newcommand{\hatT}{{\hat T}}

\newcommand{\calX}{{\cal X}}

\newcommand{\braced}[1]{{ \left\{ #1 \right\} }}

\newcommand{\floor}[1]{{\lfloor #1\rfloor}}

\newcommand{\RGreedy}{{\mbox{\sc RGreedy}}}

\newcommand{\cost}{{\mbox{\it cost}}}

\newtheorem{theorem}{Theorem}[section]

\newtheorem{lemma}[theorem]{Lemma}

\newenvironment{proof}{{\it Proof:\/}}{$\Box$\vskip 0.1in}
\newenvironment{proofsketch}{{\it Proof sketch:\/}}{$\Box$\vskip 0.1in}
                {$\Box$\vskip 0.1in}

\newenvironment{bigeqn*}{\large\begin{eqnarray*}}{\end{eqnarray*}}

\begin{document}

\title{The Reverse Greedy Algorithm for the \\
        Metric $k$-Median Problem }

\author{
       Marek Chrobak\thanks{%
       Department of Computer Science,
       University of California,
       Riverside, CA 92521.
       Email: {\tt marek@cs.ucr.edu}.
       Research supported by NSF Grant CCR-0208856.}
       \and
       Claire Kenyon\thanks{%
       Computer Science Department,
       Brown University,
       Providence, RI 02912.
       Email: {\tt claire@cs.brown.edu}. }
       \and
       Neal Young\thanks{%
       Department of Computer Science,
       University of California,
       Riverside, CA 92521.
       Email: {\tt neal@cs.ucr.edu}. }
       }

\maketitle

\begin{abstract}
  The Reverse Greedy algorithm ({\RGreedy}) for the $k$-median problem
  works as follows. It starts by placing facilities on all nodes. At
  each step, it removes a facility to minimize the total distance to
  the remaining facilities.  It stops when $k$ facilities remain.  We
  prove that, if the distance function is metric, then the
  approximation ratio of {\RGreedy} is between $\Omega(\log n/\log\log
  n)$ and $O(\log n)$.
\end{abstract}


\medskip

%%%%%%%%%%%%%%%%%%%%%%%%%%%%%%%%%%%%%%%%%%%%%%%%%%%%%%%%%%%%%%%%%%%%
%%%%%%%%%%%%%%%%%%%%%%%%%%%%%%%%%%%%%%%%%%%%%%%%%%%%%%%%%%%%%%%%%%%%
%%%%%%%%%%%%%%%%%%%%%%%%%%%%%%%%%%%%%%%%%%%%%%%%%%%%%%%%%%%%%%%%%%%%

\section{Introduction}

An instance of the \emph{metric $k$-median problem} consists of a
metric space $\calX = (X,c)$, where $X$ is a set of points and $c$ is
a \emph{distance} function (also called the \emph{cost}) that
specifies the distance $c_{xy}\ge 0$ between any pair of nodes $x,y\in
X$.  The distance function is reflexive, symmetric, and satisfies the
triangle inequality.  Given a set of points $F\subseteq X$, the cost
of $F$ is defined by $\cost(F) = \sum_{x\in X} c_{xF}$, where $c_{xF}
= \min_{f\in F} c_{xf}$ for $x\in X$.  Our objective is to find a
$k$-element set $F\subseteq X$ that minimizes $\cost(F)$.

Intuitively, we think of $F$ as a set of facilities
and of $c_{xF}$ as the cost of serving a customer at $x$ using the
facilities in $F$. Then $\cost(F)$ is the overall service
cost associated with $F$.
The $k$-element set that achieves the minimum value of $\cost(F)$
is called the \emph{$k$-median} of $\calX$.

The $k$-median problem is a classical facility location problem and
has a vast literature. Here, we review only the work most directly
related to this paper. The problem is well known to
be NP-hard, and extensive research has been done on
approximation algorithms for the metric version.
Arya \etal~\cite{arya01} show that the optimal solution can be
approximated in polynomial time within ratio $3+\epsilon$,
for any $\epsilon > 0$, and this is the smallest approximation
ratio known. Earlier, several approximation algorithms
with constant, but somewhat larger approximation ratios
appeared in the works by
Charikar~\etal~\cite{charikar99a},
Charikar and Guha~\cite{charikar99b}, 
and Jain and Vazirani~\cite{jain01}.
Jain~\etal~\cite{jain02} show
a lower bound of $1+2/e$ on the approximation ratio for this
problem (assuming P$\neq$NP).

In the {\em oblivious} version of the $k$-median problem, first studied
by Mettu and Plaxton \cite{mettu03},
the algorithm is not given $k$ in advance. Instead,
requests for additional facilities arrive over time.
When a request arrives, a new facility must be added to the
existing set. In other words, the algorithm computes
a nested sequence of facility sets
 $F_1\subset F_2 \subset \dots \subset F_n$,
where $|F_k| = k$ for all $k$.
This problem is called {\em online} median in~\cite{mettu03},
{\em incremental} median in~\cite{plaxton03}, and the analog
version for clustering is called {\em oblivious} clustering
in~\cite{CCFM97,CCFM04}. 
The algorithm presented by Mettu and Plaxton  \cite{mettu03}
guarantees that $\cost(F_k)$ approximates the optimal $k$-median
cost within a constant factor (independent of  $k$.)
They also show that in this oblivious setting no algorithm can
achieve approximation ratio better than $2-2/(n-1)$.

The naive approach to the median problem is to use
the greedy algorithm: Start with $F_0 = \emptyset$, and
at each step $k=1,\dots,n$, let $F_k = F_{k-1}\cup \braced{f_k}$,
where $f_k \in X - F_{k-1}$ is chosen so that
$\cost(F_k)$ is minimized. Clearly, this is an oblivious algorithm.
It is not difficult to show, however, that its
approximation ratio is $\Omega(n)$.

%%%%%%%%%%%%%%%

\paragraph{Reverse Greedy.}
Amos Fiat \cite{fiat02} proposed the following alternative idea.
Instead of starting with the empty set and adding facilities, start
with all nodes being facilities and remove them one by one in a greedy
fashion. More formally, Algorithm~{\RGreedy} works as follows:
Initially, let $R_n = X$. At step $k = n,n-1,\dots,2$, let $R_{k-1} =
R_k - \braced{r_k}$, where $r_k\in R_k$ is chosen so that
$\cost(R_{k-1})$ is minimized.
For the purpose of oblivious computation, the sequence of
facilities could be precomputed and then produced in order
$(r_1,r_2,\ldots,r_n)$.

Fiat \cite{fiat02} asked whether {\RGreedy} is an $O(1)$-approximation
algorithm for the metric $k$-median problem. In this note
we present a nearly tight analysis of {\RGreedy} by showing that
its approximation ratio is between $\Omega(\log n/\log\log n)$ and
$O(\log n)$. Thus, although its ratio is not constant,
{\RGreedy} performs much better than the forward greedy algorithm.

%%%%%%%%%%%%%%%%%%%%%%%%%%%%%%%%%%%%%%%%%%%%%%%%%%%%%%%%%%%%%%%%%%%%
%%%%%%%%%%%%%%%%%%%%%%%%%%%%%%%%%%%%%%%%%%%%%%%%%%%%%%%%%%%%%%%%%%%%
%%%%%%%%%%%%%%%%%%%%%%%%%%%%%%%%%%%%%%%%%%%%%%%%%%%%%%%%%%%%%%%%%%%%

\section{The Upper Bound}
\label{sec: The Upper Bound}

One crucial step of the upper bound is captured by the following lemma.

\medskip

\begin{lemma}\label{lemma: main}
Consider two subsets $R$ and $M$ of $X$. 
Denote by $Q$ the set of facilities in $R$ that
serve $M$, that is, a minimal subset
of $R$ such that $c_{\mu Q} = c_{\mu R}$ for all $\mu\in M$.
Then for every $x\in X$ we have
$c_{xQ}\leq 2c_{xM}+c_{xR}$.
\end{lemma}

\begin{proof}
For any $x\in X$, choose $r\in R$ and $\mu\in M$
that serve $x$ in $R$ and $M$, respectively. In other words,
$c_{xR} = c_{xr}$ and $c_{xM} = c_{x\mu}$.
We have $c_{\mu r}\ge c_{\mu Q}$, by the definition of $Q$.
Thus $c_{xQ} \le c_{x\mu} + c_{\mu Q}
        \le c_{x\mu} + c_{\mu r}
        \le 2c_{x\mu} + c_{xr}
        = 2c_{x\mu} + c_{xR}$.
\end{proof}

Now, fix $k$ and let $M$ be the optimal $k$-median of $\calX$.
Consider a step $j$
of {\RGreedy} (when we remove $r_j$ from $R_j$ to obtain $R_{j-1}$), for $j > k$.
Denote by $Q$ the set of facilities in $R_j$ that
serve $M$. We estimate first the incremental cost in step $j$:
\begin{eqnarray}
\cost(R_{j-1}) - \cost(R_j)
         &\le& \min_{r\in R_j\setminus Q}\cost(R_j\setminus\braced{r}) - \cost(R_j)
                \label{eqn: ub1}
                \\
         &\le& \frac{1}{|R_j\setminus Q|}
        \sum_{r\in R_j\setminus Q}[\cost(R_j\setminus \braced{r}) - \cost(R_j)]
                \\
         &\le& \frac{1}{j-k}
        \sum_{r\in R_j\setminus Q}[\cost(R_j\setminus \braced{r}) - \cost(R_j)]
                \\
         &\le& \frac{1}{j-k}
                [\cost(Q) - \cost(R_j)]
                \label{eqn: ub2}
                \\
         &\le& \frac{2}{j-k} \cost(M).
                \label{eqn: ub3}
\end{eqnarray}
The first inequality follows from the definition
of $R_{j-1}$, in the second one we estimate the minimum by the
average, and the third one follows from $|Q|\le k$.
We now justify the two remaining inequalities.

Inequality (\ref{eqn: ub2}) is related to the
the super-modularity property of the cost function.
We need to prove that 
\begin{eqnarray*}
\sum_{r\in R\setminus Q}[\cost(R\setminus\braced{r}) - \cost(R)]
                &\le& \cost(Q) - \cost(R),
\end{eqnarray*}
where $R = R_j$. To this end, we examine the contribution of each
$x\in X$ to both sides. The contribution of $x$
to the right-hand side is exactly $c_{xQ}-c_{xR}$.
On the left-hand side, the contribution of $x$ is positive
only if $c_{xQ}>c_{xR}$ and, if this is so, then $x$
contributes only to one term, namely the one for the $r\in R\setminus Q$
that serves $x$ in $R$ (that is, $c_{xr}=c_{xR}$).
Further, this contribution cannot be
greater than $c_{xQ}-c_{xR}$ because $Q\subseteq R\setminus\braced{r}$.
(Note that we do not use here any special properties of
$Q$ and $R$. This inequality holds for any $Q\subset R\subseteq X$.)

Finally, to get  (\ref{eqn: ub3}),
we apply Lemma~\ref{lemma: main} to the sets $R=R_j$, $M$, and
$Q$, and sum over all $x\in X$.

We have thus proved that 
$\cost(R_{j-1}) - \cost(R_j)\le \frac{2}{j-k} \cost(M)$.
Summing up over $j= n,n-1,\dots,k+1$, we
obtain our upper bound.

\medskip

\begin{theorem}
The approximation ratio of Algorithm~{\RGreedy}
in metric spaces is at most $2H_{n-k} = O(\log n)$.
\end{theorem}

%%%%%%%%%%%%%%%%

\section{The Lower Bound}
\label{sec: The Lower Bound}

In this section we construct an $n$-point metric space $\calX$ where,
for $k=1$, the ratio between the cost of the {\RGreedy}'s facility set
and the optimal cost is $\Omega(\log n/\log\log n)$.
(For general $k$, a lower bound of $\Omega(\log (n/k)/\log\log (n/k))$
follows easily, by simply taking $k$ copies of $\calX$.)

To simplify presentation, we allow distances between different points
in $\calX$ to be $0$. These distances can be changed to some
appropriately small $\epsilon>0$ without affecting the asymptotic
ratio. Similarly, whenever convenient, we will break the
ties in {\RGreedy} in our favor.

Let $\hatT$ be a graph that
consists of a tree $T$ with root $\rho$ and a node $\mu$
connected to all leaves of $T$. $T$ itself consists of $h$ levels
numbered $1,2,\dots,h$, with the leaves at level $1$ and the root
$\rho$ at level $h$. Each node at level $j > 1$ has $(j+1)^3$ children
in level $j-1$.

To construct $\calX$, for each node $x$ of $T$ at level $j$ we create a
cluster of $w_j = j!^3$ points (including $x$ itself) at distance
$0$ from each other. Node $\mu$ is a 1-point cluster.
All other distances are defined by shortest-path lengths in $\hatT$.

First, we show that, for $k=1$, {\RGreedy} will end up with the
facility at $\rho$. Indeed, {\RGreedy} will first remove all but one
facility from each cluster. Without loss of generality, let those
remaining facilities be located at the nodes of $\hatT$, and from now
on we will think of $w_j$ as the weight of each node in layer $j$.
At the next step, we break ties so that
{\RGreedy} will remove the facility from $\mu$. 

We claim that in any subsequent step $t$, if $j$ is the first
layer that has a facility, then {\RGreedy} has a facility on each
node of $T$ in layers $j+1,\dots,h$. To prove it, we show that this
invariant is preserved in one step. If a node $x$ in layer $j$ has
a facility then, by the invariant, this facility serves all the nodes
in the subtree $T_x$ of $T$ rooted at $x$, plus possibly $\mu$ (if
$x$ has the last facility in layer $j$.)
What facility will be removed by {\RGreedy} at this step?
The cost of removing any
facility from layers $j+1,\dots,h$ is at least $w_{j+1}$. If we
remove the facility from $x$, all the nodes served by $x$ can
switch to the parent of $x$, so the increase in cost is bounded by
the total weight of $T_x$ (possibly plus one, if $x$ serves
$\mu$.) $T_x$ has $(j+1)!^3/(i+1)!^3$ nodes in each layer $i\le j$.
So the total weight of $T_x$ is
\begin{eqnarray*}
w(T_x) &=& \sum_{i=1}^j w_i \cdot (j+1)!^3 /(i+1)!^3 
                \\
    &=& (j+1)!^3 \sum_{i=1}^j (i+1)^{-3}
            \\
    &<& (j+1)!^3
            \\
             &=& w_{j+1},
\end{eqnarray*}
where the inequality above follows from
$\sum_{i=1}^j (i+1)^{-3} \le \sum_{i=2}^\infty i^{-2} < 1$.
Thus removing $x$ increases the cost by at most $w(T_x)+1\le w_{j+1}$, so
{\RGreedy} will remove $x$ or some other node from layer $j$ in this step, as
claimed. Therefore, overall, after $n-1$ steps, {\RGreedy} will be left with
the facility at $\rho$.

By the previous paragraph, the cardinality (total weight) of $\calX$ is
$n = w(T)+1 \le (h+1)!^3$, so $h = \Omega(\log n/\log\log n)$.
The optimal cost is
\begin{eqnarray*}
\cost(\mu) &=& \sum_{i=1}^h i \cdot w_i\cdot (h+1)!^3/(i+1)!^3        
                \\
        &=&
        (h+1)!^3 \sum_{i=1}^h i(i+1)^{-3}        
                \\
        &<& (h+1)!^3 \sum_{i=2}^\infty i^{-2}        
                \\
        &<& (h+1)!^3,
\end{eqnarray*}
while the cost of {\RGreedy} is
\begin{eqnarray*}
\cost(\rho) &=&\sum_{i=1}^h (h-i) \cdot w_i\cdot (h+1)!^3/(i+1)!^3
        \\
        &=& (h+1)!^3 \sum_{i=1}^h (h-i)(i+1)^{-3}
        \\
        &\ge&
        (h-1)(h+1)!^3/8,
\end{eqnarray*}
where in the last step we estimate the sum by the first term.
Thus the ratio is $\cost(\rho)/\cost(\mu) \ge (h-1)/8 =
\Omega(\log n/\log\log n)$. 

\medskip
In the argument above we considered only the case $k=1$.
More generally, one might characterize the performance ratio of the
algorithm as a function of both $n$ and $k$. Any lower bound for $k=1$
implies a
lower bound for larger $k$ by simply taking $k$ (widely separated)
copies of the metric space. Therefore we obtain:

\begin{theorem}
The approximation ratio of Algorithm~{\RGreedy} in metric spaces
is not better than $\Omega(\log (n/k)/\log\log (n/k))$.
\end{theorem}

%%%%%%%%%%%%%%%%%%%%%%%%%%%%%%%%%%%%%%%%%%%%%%%%%%%%%%%%%%%%%%%%%%%%
%%%%%%%%%%%%%%%%%%%%%%%%%%%%%%%%%%%%%%%%%%%%%%%%%%%%%%%%%%%%%%%%%%%%
%%%%%%%%%%%%%%%%%%%%%%%%%%%%%%%%%%%%%%%%%%%%%%%%%%%%%%%%%%%%%%%%%%%%

\section{Technical Observations}

We have shown an $O(\log n)$ upper bound and an $\Omega(\log n/\log\log
n)$ lower bound on the approximation ratio of {\RGreedy} for
$k$-medians in metric spaces.  Next we make some observations about
what it might take to improve our bounds. We focus on the case $k=1$.

%%%%%%%%%%%%%%%%

\paragraph{Comments on the upper bound.}
 In the upper bound proof in 
Section~\ref{sec: The Upper Bound}  we
show that the incremental cost of {\RGreedy} when
removing $r_j$ from $R_j$ to obtain $R_{j-1}$
is at most $2\cost(\mu)/(j-1)$, where $\mu$ denotes the optimal $1$-median. 
The proof
 (inequalities (\ref{eqn: ub1}) through (\ref{eqn: ub3})) doesn't use
any information about the structure of $R_j$: it shows
that for \emph{any} set $R$ of size $j$,
\begin{eqnarray}
\min_{r} \cost(R\setminus \{r\})-\cost(R)
        &\le&
                \frac{2\cost(\mu)}{j-1}.
                        \label{eqn: general inequality}
\end{eqnarray}
Next we describe a set $R$ of size $j$ in a metric space for which
this latter bound is tight.
The metric space is defined by the following weighted graph:

%%%%%%%%%%%%%

\medskip

\begin{center}
\mbox{\ }
\medskip
\includegraphics[width=3.5in]{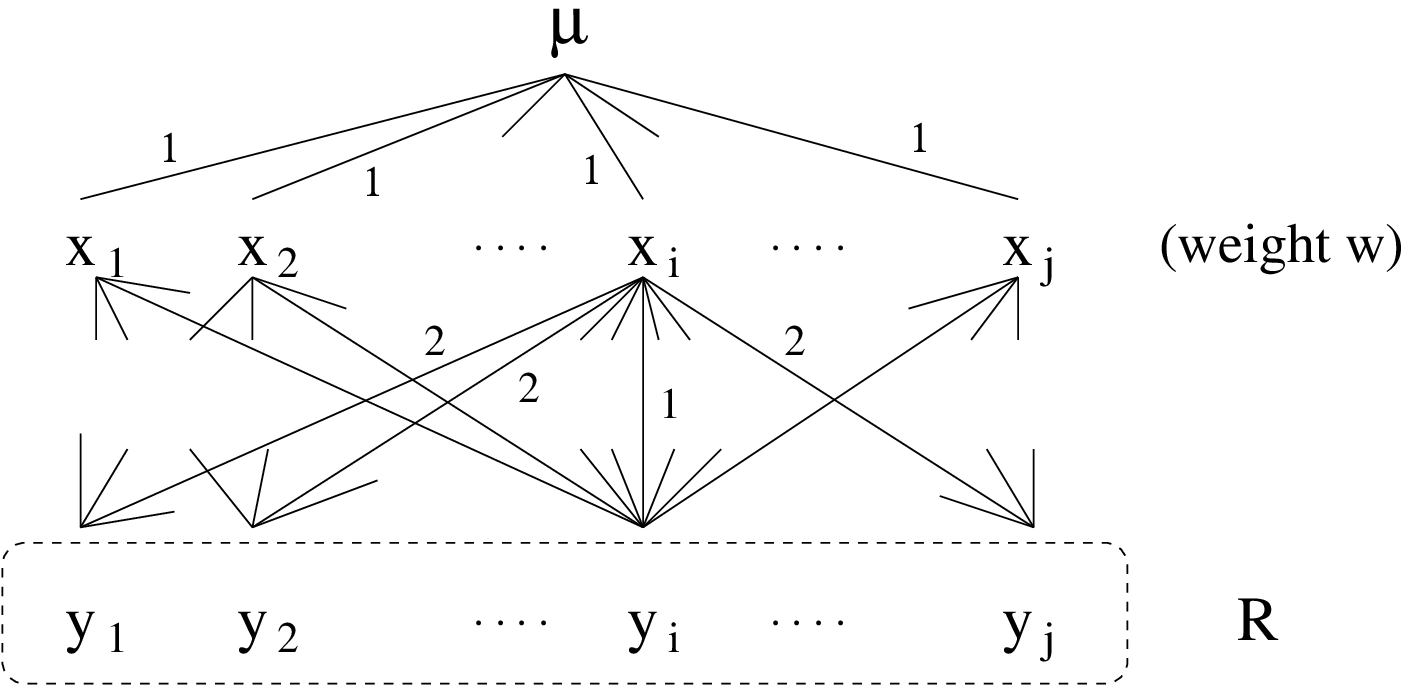}
\mbox{\ }
\medskip
\end{center}

\medskip

%%%%%%%%%%%%%

\noindent
The space has points $\mu$, $x_1,\dots,x_j$, $y_1,\dots,y_j$, where
the points $x_i$ have weights $w$, for some large integer $w$. (In
other words, each $x_i$ represents a cluster of $w$ points at distance
$0$ from each other.)  All other points have weight $1$. Point $\mu$
is connected to each $x_i$ by an edge of length $1$. Each $x_i$ is
connected to $y_i$ by an edge of length $1$, and to each $y_l$, for
$l\neq i$, by an edge of length $2$. The distances are measured along
the edges of this graph.

For $k=1$, the optimal cost is $\cost(\mu) = j(w+2)$.  Now consider $R
= \braced{y_1,\dots,y_j}$.  Removing any $y_i\in R$ increases the cost
by $w \approx \cost(\mu)/j$.  Thus, for this example, 
inequality (\ref{eqn: general inequality}) is tight,
up to a constant factor of about 2.

Of course, {\RGreedy} would not produce the particular set $R$ assumed
above for $R_j$.  Also, this example only shows a
\emph{single iteration} where the incremental cost matches 
the upper bound (\ref{eqn: general inequality}).
Nonetheless, the example demonstrates that to improve the
upper bound  it is necessary to consider some
information about the structure of $R_j$ (due to the
previous steps of {\RGreedy}).

%%%%%%%%%%%%%%%%%%%%%%%%%%%%%%%%%%%%%%%%%%%%%%%%%%%%%%%%%%%%%%%%%%%%

\paragraph{Comments on the lower bound.}
We can show that the lower-bound constructions similar to that 
in Section~\ref{sec: The Lower Bound} are unlikely to give
any improvement, in a technical sense formalized in
Lemma~\ref{lem: weird lemma}. 

Fix a metric space $\calX = (X,c)$ with $n$ points, where $n$
is a large integer.
Let $\mu$ be the $1$-median of $\calX$, and
assume (by scaling) that its cost is $\cost(\mu) = n/2$.
Let $B$ be the unit ball around $\mu$, that is,
the set of points at distance at most $1$ from $\mu$.
Note that $|B| \ge n/2$.

For $i\ge 0$, define $Z_i$ to be the points $x\in X$ such that
$i-1 < c_{x\mu} \le i$, and such that there is a time when $x$
is used by {\RGreedy} as a facility for some point in $B$.
Thus $Z_0 = \braced{\mu}$ and $Z_0\cup Z_1 = B$.
Also, for $i \le j$, let $Z_{i,j} = \cup_{l=i}^j Z_l$.

Let $h$ be the maximum index for which $Z_h\neq \emptyset$.
Define $t_j$ to be the time step when {\RGreedy} is about to remove
the last facility from $Z_{0,j}$,  and for $j\ge 7$
let $m_j$ be the number of points served by $Z_j$ at time $t_{j-6}$.
(The value of $6$ is not critical; any constant $C\ge 6$ will work,
with some minor modifications.)

%%%%%%%%%%%%%

\medskip

\begin{lemma}\label{lem: weird lemma}
Suppose that $\sum_{i=10}^h i m_i =O(n)$.
Then, for $k=1$, the approximation ratio of {\RGreedy} is
$O(\log n/\log\log n)$.
\end{lemma}

\begin{proofsketch}
We will show that $h=O(\log n/\log\log n)$. Since
the facility computed by {\RGreedy} for $k=1$
is at distance at most $h$ from $\mu$, 
this will imply the lemma, by the triangle inequality.

We first argue that $Z_i=\emptyset$ cannot happen for more than four consecutive values
of $i<h$. Indeed, $Z_0,Z_1\neq\emptyset$. Assume, towards a contradiction, 
that $Z_i\neq\emptyset$ and that
$Z_{i+1,i+4}=\emptyset$. Then at step $t_i$, 
{\RGreedy} deletes the last facility $f\in Z_{0,i}$, its
cost to serve $\mu$ increases by at least 4 and its cost to serve
$B$ increases by more than $2|B|\geq n$. Let $j>i+4$ be such that $Z_j\neq \emptyset$.
By Lemma~\ref{lemma: main},
deleting a facility $f'\in Z_j$ at time $t_i$ would increase the cost
by at most $2\cost(\mu) \le n$, hence
less than the cost of deleting $f$ at time $t_i$ --
contradicting the definition of  {\RGreedy}.
 
Now, consider any $i \le h - 9$.
It is easy to see that over all steps $t_i,t_i+1,..,t_{i+3}$,
{\RGreedy}'s cost to serve $B$ increases by at least
$|B|\ge n/2$, while, by the triangle inequality,
all facilities that serve $B$ at steps $t_{i+1},t_{i+1}+1,...,t_{i+3}$ are
in $Z_{i+1,i+5}$.
Thus, there exists a $t\in [t_i,t_{i+3}]$ such that at step $t$,
{\RGreedy} deletes a facility $f$ and pays an incremental cost of at least
$(n/2)/(1+|Z_{i+1,i+5}|)$.

Suppose $Z_{i+9}\neq\emptyset$.
Since $t\le t_{i+3}$, the facilities in $Z_{i+9}$ serve at
most $m_j$ clients. Therefore,
at step $t$, deleting \emph{all} facilities in $Z_{i+9}$ and 
serving their clients using a remaining facility from 
$Z_{i,i+3}$ would have
increased the cost by $O(im_{i+9})$, by the triangle inequality.
So there exists a facility $f'$ in
$Z_{i+9}$ whose deletion at step $t$ would have increased the cost by
$O(im_{i+9}/|Z_{i+9}|)$.  Since at time $t$ {\RGreedy} prefers to
delete $f$ rather than $f'$, we have 
\[
  (n/2)/(1+|Z_{i+1,i+5}|) \,=\, O(im_{i+9}/|Z_{i+9}|).
\]
Rewriting and summing the above over $i$
(including now those $i$ for which $Z_{i+9}$ is empty),
\begin{equation}\label{eq: lemma4.1}
  \sum_{i=1}^{h-9} \frac{|Z_{i+9}|}%
  {1+|Z_{i+1,i+5}|}
  \,=\, O\Big(\frac{1}{n} \sum_{i=1}^{h-9} i m_{i+9}\Big)
  \,=\, O\Big(\frac{1}{n} \sum_{i=10}^h i m_{i}\Big)
\;\leq \; A,
\end{equation}
for some constant $A$.

The intuition is that for this sum to be bounded by a constant,
the cardinalities $|Z_i|$ must rapidly decrease (except for
some small number of abnormalities) and $h$ cannot be
too large. To get a good estimate, let $y_i = |Z_{8i+1,8i+8}|$,
for $i = 1,\dots,\floor{h/8}-1$. Then, 
\begin{eqnarray*}
\sum_{i=1}^{\floor{h/8}-2}\frac{ y_{i+1} }{ y_i + y_{i+1} }
         \;=\; \sum_{i=1}^{\floor{h/8}-2}
                 \sum_{j=8i+1}^{8i+8} \frac{ | Z_{j+8}| }{ |Z_{8i+1,8i+16}| }
        \;\le\;
\sum_{i=1}^{\floor{h/8}-2}
       \sum_{j=8i+1}^{8i+8} \frac{ | Z_{j+8}| }{ 1+ |Z_{j,j+4}| }
        \;\le\; A,
\end{eqnarray*}
where the next-to-last inequality holds because $1+|Z_{j,j+4}|\le |Z_{8i+1,8i+16}|$
for all $j = 8i+1,...,8i+12$. (Here, again, we use the fact that at most four
consecutive $Z_l$'s can be zero.)

Now let $q_i = y_{i+1}/y_i$ for all $i= 1,\dots,\floor{h/8}-2$.
We have $\sum_{i=1}^{\floor{h/8}-2} q_i/(1+q_i) \le A$.
Therefore $q_i\le 1$ for all except at most $2A$ $i$'s.
So there are $m$ and $g\ge (\floor{h/8}-2)/(2A)$ such that
$q_i\le 1$ for all $i = m,...,m+g-1$. For those $i$'s we get
\begin{eqnarray*}
\sum_{i=m}^{m+g-1} q_i
         \;\le\; 2\cdot \sum_{i=m}^{m+g-1} \frac{q_i}{1+q_i}
         \;=\; 2\cdot \sum_{i=m}^{m+g-1} \frac{y_{i+1}}{y_i+y_{i+1}}
         \;\le\; 2A.
\end{eqnarray*}
Let $\sum_{i=m}^{m+g-1} q_i = B \le 2A$. Then $\prod_{i=m}^{m+q-1} q_i$
is maximized when all $q_i$ are equal to $B/g$, and therefore
\begin{eqnarray*}
\frac{1}{n} \;\le\; \frac{y_{m+g}}{y_m} 
        \;=\; \prod_{i=m}^{m+g-1} q_i
        \;\le\; (B/g)^g.
\end{eqnarray*}
Thus $(g/B)^g \le n$, and we obtain
$h = O(g) = O(\log n/\log\log n)$, completing the proof.
\end{proofsketch}

Note that assumption of the lemma holds for the metric space used in
Section~\ref{sec: The Lower Bound}.
There, each set $Z_i$, for $i=1,...,h$, consists of the nodes in $T$ at level $i$,
and $m_i = (h+1)!^3/(i+1)^3$ is the total weight of level $i$ so, indeed,
$\sum_{i=1}^h im_i = O(h!^3) = O(n)$.
The lemma suggests that in order
to improve the lower bound, one would need to design an example where
at every time $t_i$, the facilities serving nodes at distance at most $i$
from $\mu$ are  distributed more or less
uniformly across the remaining facilities.

%%%%%%%%%%%%%%%%%%%%%%%%%%%%%%
%%%%%%%%%%%%%%%%%%%%%%%%%%%%%%
%%%%%%%%%%%%%%%%%%%%%%%%%%%%%%

\bigskip
\paragraph{Acknowledgments.} We would like to thank
Amos Fiat, Christoph D{\"u}rr, Jason Hartline, Anna Karlin,
and John Noga for useful discussions.

%%%%%%%%%%%%%%%%%%%%%%%%%%%%%%%%%%%%%%%%%%%%%%%%%%%%%%%%%%%%%%%%%%%%
%%%%%%%%%%%%%%%%%%%%%%%%%%%%%%%%%%%%%%%%%%%%%%%%%%%%%%%%%%%%%%%%%%%%
%%%%%%%%%%%%%%%%%%%%%%%%%%%%%%%%%%%%%%%%%%%%%%%%%%%%%%%%%%%%%%%%%%%%

\end{document}